\documentclass[seceq]{ptptex}
\usepackage{graphicx}
\usepackage{wrapft}

\notypesetlogo
\title{Distance-Redshift Relation in a Realistic Inhomogeneous Universe}
\author{Tomohiro \textsc{Okamura}\footnote{E-mail address: t-okamura@astr.tohoku.ac.jp} 
and Toshifumi \textsc{Futamase}\footnote{E-mail address: tof@astr.tohoku.ac.jp}
}

\inst{Astronomical Institute, Tohoku University, \\
Sendai 980-8578, Japan \\}

\abst{
We investigate the distance-redshift relation in a realistic inhomogeneous universe 
where the mass distribution is described by the mass function of Sheth and Tormen.  
It is found that the derived distance deviates systematically from the standard distance up to 10\% 
depending on the choice of the lowest halo mass in which baryonic matter condensed to form luminous object such as galaxies. 
Remarkably the derived distance is well approximated by the Dyer-Roeder distance 
if we choose the clumpiness parameter $\alpha$ calculated by our model. 
We also discuss the effect of inhomogeneities in the determination of dark energy parameter 
in the supernovae observation, and find that this effect must be taken into account for the future high redshift 
supernovae observation. 
}

\begin{document}
\maketitle

\section{Introduction}

It is needless to say that the distance-redshift relation has a fundamental importance in the observational cosmology. 
The distance formula is usually derived by assuming that the universe is homogeneous and isotropic. 
This assumption has been supported by the observations of large scale galaxy survey and cosmic microwave background radiation over the scale of about 100Mpc.
The derived formula is called the standard distance or the filled beam distance. 
However the light rays feel not the averaged geometry but local inhomogeneous geometry, and thus it is reasonable to expect that the local inhomogeneous matter distribution does have some effect on the distance-redshift relation. 
There have been various attempts to derive the distance-redshift relation taking the effect of clumpiness of matter distribution into account.\cite{rf:1,rf:2,rf:3,rf:4,rf:5,rf:6}
Most well known is the so-called Dyer-Roeder (DR) distance which is derived under the assumption that the light rays propagate the vacuum region along their entire paths. 
This is a reasonable assumption in a sense since we do not observe the distant light sources inside galaxy this side of the source.    
In fact there are some studies supporting the validity of DR distance in a simplified situation where dark matter distribution is not taken into account. 
However it is now widely accepted that dark matter is an important constitution of the universe and plays an essential role 
in the structure formation. Thus the correct distance-redshift relation must take the distribution of dark matter into account. 
The purpose of this paper is to construct such a distance-redshift relation. 

In this paper we describe the dark matter distribution in terms of the mass function proposed by Sheth and Tormen. 
Moreover we assume that the light rays from the distant light sources cannot propagate through the mass concentration above 
a minimum mass. 
This is supposed to mean that the the baryonic matter collapses to form luminous objects such as galaxies in these mass concentrations and prevent the distant light. 
Thus the light rays propagate statistically underdense regions and it is expected that this will cause systematic deviation from the standard distance-redshift relation. 

The structure of this paper is as follows.  We first describe our model of inhomogeneities and explain our assumption in the light propagation. In Section 3 we then apply the above model to the distance-redshift relation given by Futamase and Sasaki which applies realistic inhomogeneous universe. We will find that the relation systematically predicts larger distance than the standard distance 
depending on the lowest mass introduced in Section 2.  
We will find in Section 4 that the so-called Dyer-Roeder distance gives a very good approximation to the derived distance if we use 
the clumpiness parameter $\alpha$ calculated by our model of inhomogeneities. 
We apply our result to the determination of the parameter of equation of state in the dark energy in Section 5.
Finally Section 6 will be devoted to some discussion.  

For fiducial cosmology, we adopt in this paper a totally flat universe model with $h=0.7, \Omega_{\rm de}=0.72, \Omega_{\rm m}=1-\Omega_{\rm de}=0.28, \Omega_{\rm b}=0.046, w_0=-1, w_a=0, \sigma_8=0.82, n_S=0.96$.

\section{Model of mass distribution  and the light propagation}

In a real universe, matter distribution is very clumpy and we model this clumpy universe by halo model.\cite{rf:7} \ 
The comoving number density of collapsed object (halo) which has mass $m$ in the mean comoving mass density $\bar{\rho}_{\rm m}$ at redshift $z$, $n(m,z)$, is described by 
\begin{eqnarray}
\frac{m^2n(m,z)}{\bar{\rho}_{\rm m}}\frac{dm}{m}=\nu f(\nu)\frac{d\nu}{\nu}
\end{eqnarray}
The height of density peak, $\nu$, is defined as
\begin{eqnarray}
\nu=\frac{{\delta_{\rm c}}^2(z)}{\sigma^2(m,z)} 
\end{eqnarray}
where $\delta_{\rm c}(z)\simeq1.686\Omega_{\rm m}^{0.0055}(z)$ is the overdensity of a spherical region that collapses at $z$,\cite{rf:8} \ $\Omega_{\rm m}(z)$ is the ratio of matter density to critical density at $z$.
$\sigma(m,z)$ is the rms value of the density contrast when smoothed with a tophat filter of scale $R=(3m/4\pi \bar{\rho}_{\rm m})^{1/3}$, extrapolated using linear theory to $z$ :
\begin{eqnarray}
\sigma^2(m,z)=\int\frac{dk}{k}\frac{k^3P^{\rm lin}(k,z)}{2\pi^2}|W(kR)|^2
\end{eqnarray}
where $P^{\rm lin}(k,z)$ is the linear matter power spectrum at $z$ and $W(x)=(3/x^3)(\sin{x}-x\cos{x})$ is tophat filter in fourier space.
To predict the halo number density, we use the Sheth and Tormen's mass function \cite{rf:9}
\begin{eqnarray}
\nu f(\nu)=A\left(1+\frac{1}{\nu'^p}\right)\left(\frac{\nu'}{2}\right)^{1/2}\frac{e^{-\nu'/2}}{\sqrt{\pi}}
\end{eqnarray}
with $\nu'=0.75\nu, p=0.3$ and normalization constant $A$ is chosen to satisfy the constraint $\int f(\nu)d\nu=1$.
When we  calculate the linear matter power spectrum, we use the BBKS transfer function, \cite{rf:10} \ the shape parameter from Sugiyama\cite{rf:11} \ and the growth suppression rate, $g(z)=D_{\rm lin}(z)/a(z)$, which  obey the following differential equation
\begin{eqnarray}
2\frac{d^2g}{d\ln{a^2}}+[5-3w(z)\Omega_{\rm de}(z)]\frac{dg}{d\ln{a}}+3[1-w(z)]\Omega_{\rm de}(z)g(z)=0
\end{eqnarray}
where $a(z)=1/(1+z)$ is the scale factor, $\Omega_{\rm de}(z)$ is the ratio of dark energy density to critical density at $z$ and $w(z)$ is the equation of state of dark energy.
Hereafter we parametrize $w(z)=w_0+w_a(1-1/(1+z))$.

In the standard structure formation theory, halo which has sufficient mass can attract baryonic matter and forms luminous objects 
such as galaxies. So if we chose this mass scale appropriately, we can predict the number of halo which contains luminous objects and mass of these halos.
If these halos exit between observer and sources, light rays from sources have the possibility to be prevented by these halos or extincted by dust associated with these halos.
In this paper, we consider this possibility to its maximum.
For very large halo, like a cluster of galaxies with very large $M/L\sim 200$, this assumption may not be appropriate, because of large fraction of the 'transparent' matter.
But the effect of these large halo is very small in the distance as we will see later.

Based on the above assumption, we can define the effective comoving matter density (or smooth distributed matter density) which light right ray can passes through    
\begin{align}
\rho_{\rm eff}(z)\equiv\bar{\rho}_{\rm m}-\int_{M_{\rm min}}^{M_{\rm max}}dm\,mn(m,z)
\end{align}

\begin{figure} 
\centering
\includegraphics[angle=-90,width=11cm]{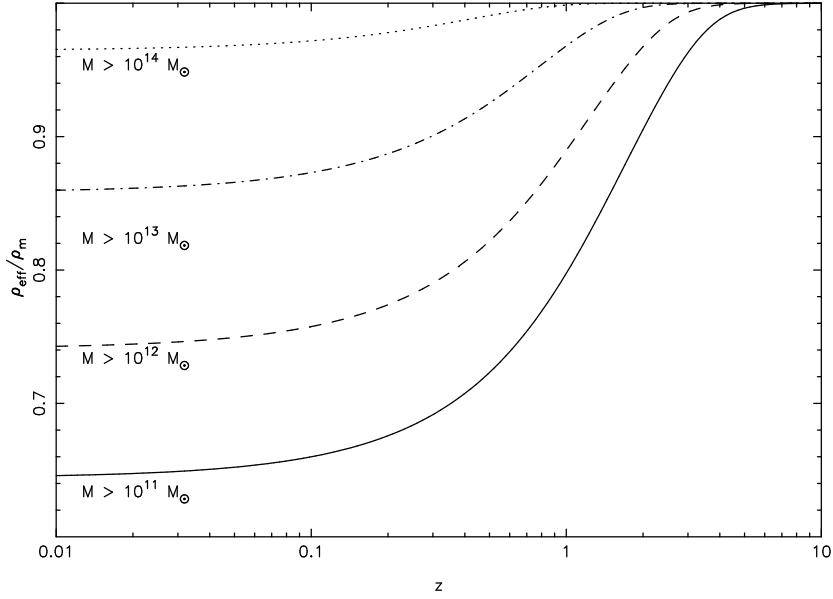}
\caption{
\label{fig:alpha}
The ratio of effective comoving matter density to total comoving matter density as a function of redshift for various mass range:
lowest mass is selected at $10^{11}M_{\odot}$(solid), $10^{12}M_{\odot}$(dashed), $10^{13}M_{\odot}$(dot-dashed) and $10^{14}M_{\odot}$(dotted).
} 
\end{figure}

$M_{\rm min}$ is a minimum halo mass in which baryonic matter condensed to form luminous objects and is our parameter. 
Figure 1 shows the ratio of effective comoving mass density to total comoving mass density with various choice of the minimum mass.
One can see that very large halos ($M_{\rm min}= 10^{14}M_{\odot}$) have little effect on the effective density. This is because 
they have large masses but is very small. On the other hand, halos with mass range $10^{11}M_{\odot}< M < 10^{13}M_{\odot}$ 
contributes most effectively because $m\times n(m,z)$ is the largest around $10^{12}M_{\odot}$ at $z=0$.  
As redshift gets larger, an inhomogeneity becomes small and $\rho_{\rm eff}$ approaches to $\bar{\rho}_{\rm m}$ at $z \sim O(1)$.
In these calculation, we chose highest masses infinity since the contribution by masses much larger than $10^{14}M_{\odot}$ may be 
totally neglected.

\section{Distance-redshift relation}
In a homogeneous flat FRW universe, the angular diameter distance is defined as \cite{rf:12}
\begin{eqnarray}
d^{\rm FRW}_{\rm A}(z)=\frac{\chi(z)}{1+z}
\end{eqnarray}
$\chi(z)$ is the comoving distance in a homogeneous universe, which is an integral over redshift of inverse Hubble parameter $H(z)$
\begin{align}
\chi(z)=\int_0^{z}\frac{dz'}{H(z')}
\end{align}

Using the optical scalar equations \cite{rf:13} and the perturbation theory, Futamase and Sasaki showed that in a realistic inhomogeneous universe,  the angular diameter distance is modified as \cite{rf:14,rf:15,rf:16}
\begin{align}
\delta_{d_{\rm A}}(z,\hat{n})\equiv\frac{\delta d_{\rm A}^{\rm FRW}(z,\hat{n})}{d^{\rm FRW}_{\rm A}(z)}=&v_{\rm s}^i\hat{n}_i-\frac{1}{\chi_{\rm s}}\left[\frac{1}{aH}\right]_{\rm s}(v_{\rm s}^i\hat{n}_i-v_{\rm o}^i\hat{n}_i) \nonumber \\
&-\frac{3H_0^2\Omega_{\rm m}}{2}\int_0^{\chi_s}d\chi\frac{(\chi_s-\chi)\chi}{\chi_s}(1+z')\delta_{\rm m}(z',\hat{n})
\end{align}
where $\hat{n}$ is the source direction, $\chi_s$ is the source comoving distance, $v_s$, $v_o$ is the source and observer peculiar velocity respectively and $\delta_{\rm m}$ is the matter overdensity.
The first line is so called the doppler term because of local peculiar velocity, which changes the redshift of sources relative to observer and the solid angle of observer.
The second line is so called the lensing term because of inhomogeneity of line-of-sight matter distribution
(more detailed physical description is described in Ref.~\citen{rf:17}).
Then, we define 
\begin{align}
d^{\rm FS}_{\rm A}(z)\equiv d^{\rm FRW}_{\rm A}(z)\left(1+\left\langle\delta_{d_{\rm A}}(z,\hat{n})\right\rangle_b\right)
\end{align}
$\langle\dots\rangle$ means ensemble average and subscript '$b$' denotes that we can only average using observed sources.
Of course, if we can average all perturbed density field, this distance return to $d^{\rm FRW}_{\rm A}$.
But under our assumption discussed in \S 2, average of $\delta_{\rm m}$ doesn't vanish
\begin{align}
\left\langle\delta_{\rm m}(z,\hat{n})\right\rangle_b=\frac{\rho_{\rm eff}}{\bar{\rho}_{\rm m}}-1
\end{align}
This means that light ray which can arrive us feels limited overdensity and tends to pass through underdense region, so $d^{\rm FS}_{\rm A}$ doesn't equal to $d^{\rm FRW}_{\rm A}$.
From Fig.1, $\rho_{\rm eff}/\bar{\rho}_{\rm m}$ is always smaller than 1, so we expect that $d^{\rm FS}_{\rm A}$ is always larger than $d^{\rm FRW}_{\rm A}$. 
 
Figure 2 shows the fractional differences of $d^{\rm FS}_{\rm A}$ and $d^{\rm FRW}_{\rm A}$ as a function of redshift for various mass range.
As expected above, $d^{\rm FS}_{\rm A}$ is always larger than $d^{\rm FRW}_{\rm A}$.
The lensing weight function, $(\chi_s-\chi)\chi/\chi_s$, makes this effect continue to grow by higher redshift compared to $\left\langle\delta_{\rm m}\right\rangle_b$.
Very large halos ($M>10^{14}M_{\odot}$) change distance a little ($\lesssim 0.1\%$) and we can ignore.
But middle large halos change distance largely up to 10\% and we can't ignore at high redshift apparently.
At $z \sim 1$, this effect look mildly small, but in searching for dark energy property precisely this change may make some bias to dark energy parameter, we will discuss in \S 5.

\begin{figure} 
\centering
\includegraphics[angle=-90,width=11cm]{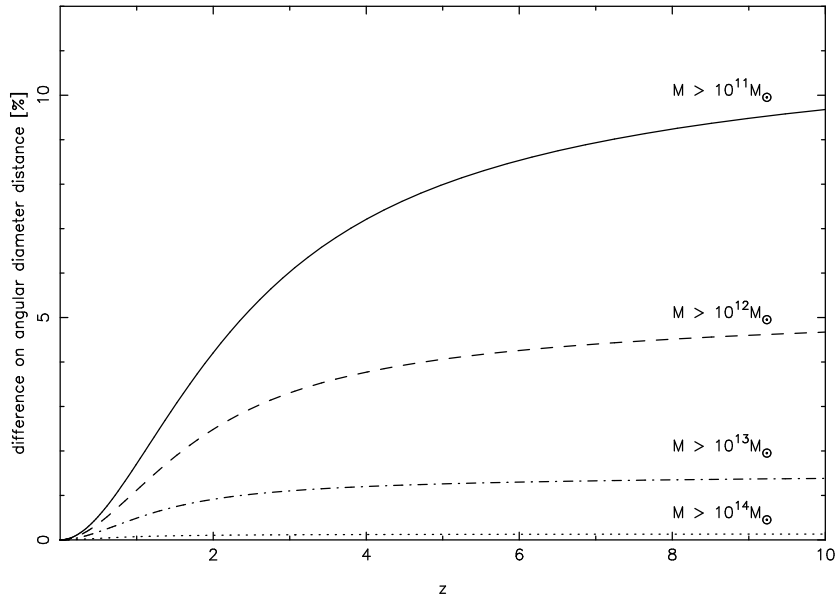}
\caption{
\label{fig:fs}
The fractional differences on angular diameter distance between $d^{\rm FS}_{\rm A}$ and $d^{\rm FRW}_{\rm A}$ as a function of redshift, which is defined as $(d^{\rm FS}_{\rm A}/d^{\rm FRW}_{\rm A}-1$).
They are defined in text respectively.
The mean of curves is the same as Fig. 1.
} 
\end{figure}

\section{Dyer-Roeder distance}
There have been many discussions on a possible effect of local inhomogeneities in real universe on the distance-redshift relation 
starting from the discussion by Zel'dovich.\cite{rf:18} \ 
For example, Dyer and Roeder derived the distance formula on the assumption that a certain fraction, the constant 
clumpiness parameter $\alpha$, of all matter is distributed uniformly whereas the rest is clumped into galaxies, and that light rays 
travel well away from all clumps of matter, feeling only the effect of the fraction $\alpha$ of all matter.\cite{rf:19} \ 
In Dyer-Roeder formalism, angular diameter distance, $d_{\rm A}^{\rm DR}(z)$, obeys the following differential equation

\begin{equation}
(1+z)^2{\cal F}\frac{d^2d_{\rm A}^{\rm DR}}{dz^2}+(1+z){\cal G}\frac{d\,d_{\rm A}^{\rm DR}}{dz}+{\cal H}d^{\rm DR}_{\rm A}=0
\end{equation}
which satisfies the boundary conditions:
\begin{subequations}
\begin{align}
d^{\rm DR}_{\rm A}(0)&=0 \\
\left.\frac{d\,d_{\rm A}^{\rm DR}}{dz}\right|_{z=0}&=\frac{1}{H_0}
\end{align}
\end{subequations}
where
\begin{subequations}
\begin{align}
{\cal F}=&H^2(z) \\
{\cal G}=&(1+z)H(z)\frac{dH}{dz}+2H^2(z) \\
{\cal H}=&\frac{3\alpha(z)}{2}\Omega_{\rm m}(1+z)^3+\frac{3(w(z)+1)}{2}(1-\Omega_{\rm m})\exp{\left[3\int_0^z \frac{dz'}{1+z'}(w(z')+1)\right]}
\end{align}
\end{subequations}
For our fiducial model, $w(z)=-1$, the second term of ${\cal H}$ vanishes.
Of course, if $\alpha=1$, ${\cal H}$ becomes $(1+z)H(z)dH/dz$ and return to the FRW distance.
In our inhomogeneous model, the clumpiness parameter $\alpha$  is equal to $\rho_{\rm eff}/\bar{\rho}_{\rm m}$ as plotted in Fig 1.
  
Figure 3 shows the fractional differences of $d^{\rm DR}_{\rm A}$ and $d^{\rm FRW}_{\rm A}$ as a function of redshift for various mass range.
For $\alpha < 1$, light beam is defocused and $d^{\rm DR}_{\rm A}$ is allways larger than $d^{\rm FRW}_{\rm A}$.
The figure shows that it is remarkably close to $d^{\rm FS}_{\rm A}$ with deviation $\lesssim 0.3\%$.

\begin{figure} 
\centering
\includegraphics[angle=-90,width=11cm]{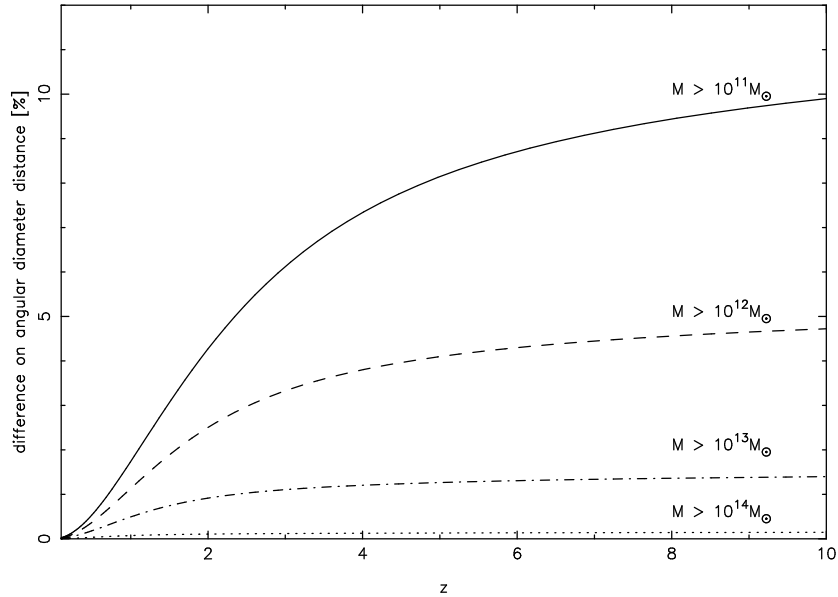}
\caption{
\label{fig:dr}
The fractional differences on angular diameter distance between $d^{\rm DR}_{\rm A}$ and $d^{\rm FRW}_{\rm A}$ as a function of redshift, which is defined as $(d^{\rm DR}_{\rm A}/d^{\rm FRW}_{\rm A}-1$).
They are defined in text respectively.
The mean of curves is the same as Fig. 1.
} 
\end{figure}

\section{Parameter biases for future SNIa survey}
In this section, we consider how this difference on distance affects the estimation of cosmological parameters based 
on the observational probe, especially Type Ia supernova (SNIa) survey.
For SNIa survey, observable quantity is the apparent magnitude $m(z)$ which relates the luminosity distance
\begin{align}
m(z)=5\log_{10}{d_{\rm L}(z)}+M=\frac{5}{\ln{10}}\ln{d_{\rm L}(z)}+M
\end{align}
where $M$ is the magnitude zero-point.
Rescaling $d_{\rm L}$, we can subtract Hubble constant $H_0$ uncertainty by the redefinition of $M$.
So our interest parameter is $\theta_i=\{\Omega_{\rm de},w_0,w_a,M\}$.
Recalling the relation between the luminosity distance and the angular diameter distance, $d_{\rm L}=(1+z)^2d_{\rm A}$, we can write difference on apparent magnitude in our model
\begin{align}
\delta m(z)=\frac{5}{\ln{10}}\ln{(1+\left\langle\delta_{d_{\rm A}}(z,\hat{n})\right\rangle_b)}
\end{align}

SNIa survey has some uncertainty which comes from not only astrophysical, but also cosmological ones.\cite{rf:17,rf:20} \ 
In this paper, we only consider the intrinsic dispersion of magnitude $\sigma_{\rm m}$.
If the noise covariance consists of only $\sigma_{\rm m}$, bias to parameter vector $\theta_i$ is described as \cite{rf:21}
\begin{eqnarray}
\delta\theta_i={F^{-1}}_{ij}\frac{1}{\sigma^2_{\rm m}}\int dzN(z)\delta m(z)\frac{\partial m}{\partial \theta_j}
\end{eqnarray}
where $F_{ij}$ is the Fisher matrix of fiducial survey :
\begin{align}
F_{ij}=\frac{1}{\sigma^2_{\rm m}}\int dzN(z)\frac{\partial m}{\partial \theta_i}\frac{\partial m}{\partial \theta_j}
\end{align}
and $N(z)$ is the number of SNIa at $z$.
We consider SNAP like survey and use survey parameters written in Kim et al.\cite{rf:21} \  
Substituting (5.4) into (5.3), $\sigma_{\rm m}$ is cancelled out.
So we don't have to chose $\sigma_{\rm m}$ in bias calculation and our result is not affected by $\sigma_{\rm m}$ choosing (If we consider other uncertainty, this does not hold anymore). 

Firstly, we consider bias to parameter vector $\theta_i=\{\Omega_{\rm de},w_0,M\}$ (constant dark energy model) to 
understand physical meaning.
Table 1 shows biases to parameter estimation for various mass ranges with and without prior $\sigma(\Omega_{\rm de})=0.03$ which is anticipated from other cosmological probes.
Intuitively, to make distance larger, $\Omega_{\rm de}$ becomes larger or $w_0$ becomes smaller.
Table 1 makes this expectation apparent.
Without prior, $\Omega_{\rm de}$ becomes larger, but $w_0$ becomes larger because of degeneration.
With prior, $\Omega_{\rm de}$ becomes larger slightly because of prior limitation and $w_0$ becomes smaller for compensate larger distance. It is found that the effect of local inhomogeneities affects the parameter $w_0$ of the order of 0.1, thus 
this effect must be taken into account in the future supernovae survey. 

\begin{table}
\caption{The biases to parameter vector $\theta_i=\{\Omega_{\rm de},w_0,M\}$ for various mass ranges with and without prior $\sigma(\Omega_{\rm de})=0.03$.
}
\label{table:1}
\begin{center}
\begin{tabular}{ccc}\hline \hline
mass range & $\delta\Omega_{\rm de}$(w.o/w prior) & $\delta w_0$(w.o/w prior) \\ \hline
$M > 10^{11}M_{\odot}$  &
$7.4\times10^{-2}/8.8\times10^{-3}$ &
$1.6\times10^{-1}/-1.8\times10^{-1}$ \\ 

$M > 10^{12}M_{\odot}$  &
$4.4\times10^{-2}/5.2\times10^{-3}$ &
$8.2\times10^{-2}/-1.2\times10^{-1}$  \\
 
$M > 10^{13}M_{\odot}$  &
$1.5\times10^{-2}/1.8\times10^{-3}$ &
$2.1\times10^{-2}/-4.9\times10^{-2}$  \\

$M > 10^{14}M_{\odot}$  &
$1.3\times10^{-3}/1.5\times10^{-4}$ &
$-4.3\times10^{-4}/-6.4\times10^{-3}$   \\

$10^{11}M_{\odot} < M < 10^{13}M_{\odot}$ &
$5.9\times10^{-2}/7.1\times10^{-3}$ &
$1.4\times10^{-1}/-1.4\times10^{-1}$  \\ \hline
\end{tabular}
\end{center}
\end{table}

Next, we consider bias to parameter vector $\theta_i=\{\Omega_{\rm de},w_0,w_a,M\}$ (time variable dark energy model).
Table 2 shows biases to parameter estimation for various mass ranges with and without prior $\sigma(\Omega_{\rm de})=0.03$. 
Similar to above, to make distance larger, $\Omega_{\rm de}$ becomes larger or $w_0$ becomes smaller or $w_a$ becomes smaller.
Without prior, $\delta\Omega_{\rm de}$ and $\delta w_0$ is always positive but $\delta w_a$ is positive at low mass model and negative at high mass model.
With prior, $\delta\Omega_{\rm de}$ and $\delta w_0$ are always positive and $\delta w_a$ is always negative.
We can understand behavior of $\delta\Omega_{\rm de}$ and $\delta w_a$ similarly to above.
But in spite of mass range and prior, $\delta w_0$ is always positive. 

\begin{table}
\caption{The biases to parameter vector $\theta_i=\{\Omega_{\rm de},w_0,w_a,M\}$ for various mass ranges with and without prior $\sigma(\Omega_{\rm de})=0.03$.
}
\label{table:2}
\begin{center}
\begin{tabular}{cccc}\hline \hline
mass range & 
$\delta\Omega_{\rm de}$(w.o/w prior) &
$\delta w_0$(w.o/w prior) & 
$\delta w_a$(w.o/w prior) \\ \hline

$M > 10^{11}M_{\odot}$  &
$1.0\times10^{-1}/3.4\times10^{-4}$ &
$1.1\times10^{-1}/2.6\times10^{-1}$ &
$5.5\times10^{-1}/-1.4$ \\

$M > 10^{12}M_{\odot}$  &
$5.3\times10^{-2}/1.8\times10^{-4}$ &
$6.7\times10^{-2}/1.4\times10^{-1}$ & 
$1.8\times10^{-1}/-8.3\times10^{-1}$ \\
 
$M > 10^{13}M_{\odot}$  &
$1.4\times10^{-2}/4.5\times10^{-5}$ &
$2.3\times10^{-2}/4.3\times10^{-2}$ &
$-3.3\times10^{-2}/-3.0\times10^{-1}$ \\

$M > 10^{14}M_{\odot}$  &
$3.9\times10^{-4}/1.3\times10^{-6}$ &
$1.0\times10^{-3}/1.6\times10^{-3}$ &
$-1.8\times10^{-2}/-2.6\times10^{-2}$  \\

$10^{11}M_{\odot} < M < 10^{13}M_{\odot}$ &
$6.9\times10^{-2}/2.9\times10^{-4}$ &
$8.7\times10^{-2}/2.2\times10^{-1}$ &
$3.1\times10^{-1}/-1.1$ \\ \hline
\end{tabular}
\end{center}
\end{table}

We also consider influence on the current SNIa data, especially 'UNION' data.\cite{rf:22} \footnote{SNIa data is available from http://supernova.lbl.gov/Union/} \ 
Doing similar calculation above (we use each $\sigma_{\rm m}$ available from above URL), we find that bias to parameter vector $\theta_i=\{\Omega_{\rm de}, w_0, M\}$ with prior $\sigma(\Omega_{\rm de})=0.03$ in mass model $10^{11}M_{\odot} < M < 10^{13}M_{\odot}$ is $\delta\Omega_{\rm de}=1.1\times 10^{-5}$ and $\delta w_0=-4.1\times 10^{-2}$.
Thus this effect is negligible compared to statistical error in the current SNIa data because of low redshift SNIa.

\section{Conclusion and discussion}
In this paper we investigate the distance-redshift relation in a realistic inhomogeneous universe 
where the mass distribution is described by the mass function of Sheth and Tormen.  
It is found that the derived distance deviates systematically from the standard distance up to 10\% depending on the choice of the lowest mass. 
Remarkably the derived distance is well approximated by the Dyer-Roeder distance 
if we choose the clumpiness parameter $\alpha$ calculated by our model. 
We also discuss the effect of inhomogeneities in the determination of dark energy parameter 
in the supernovae observation. The effect behaves like dark energy and thus is relevant in dark energy survey.  
Although it does not change the result of current SNIa survey, it should be seriously taken into account future 
SNIa survey since it aims at deciding the dark energy parameter in a few \% statistical error.

In the present formalism we have not considered the dispersion of the distance-redshift relation. 
It will be very interesting to consider the dispersion for more realistic interpretation of the observed data. 
The dispersion itself has important meaning in cosmology because it carries the information of large scale structure as well as dark energy property.
The dispersion depends on the spatial and temporal distribution of halos and we would like to consider this problem in future work.  

We have applied our distance model to SNIa survey in this paper. 
It is also interesting to find other applications such as the evolution of the luminosity function of quasar 
based on our distance model.

\section*{acknowledgments}
We would like to thank for Y. Okura and M. Kilbinger for useful discussions.
This work is also supported in part by a Grants-in-Aid for Scientific Research from JSPS (Nos. 18072001, 20540245 for TF) as well as by Core-to-Core Program "International Research Network for Dark Energy".

\end{document}